\documentclass{elsart}
\usepackage{graphicx}
\journal{Physics Letters A}

\begin{document}

\begin{frontmatter}

\title{Integrable discretizations of a two-dimensional Hamiltonian system
with a quartic potential}

\author{Bao-Feng Feng and Ken-ichi Maruno}

\address{Department of Mathematics, The University of Texas-Pan
American, Edinburg, Texas, USA 78541-2999}
\date{\today}


\begin{abstract}
In this paper, we propose integrable discretizations of a
two-dimensional Hamiltonian system with quartic potentials. Using
either the method of
separation of variables or the method based on bilinear forms,
we construct the corresponding integrable
mappings for the first three among four integrable cases.
\end{abstract}

\begin{keyword}
Integrable mappings, Hamiltonian system, Separation of variables
\PACS 05.45.Yv \sep 45.05.+x \sep 47.10.Df
\end{keyword}
\end{frontmatter}

\section{Introduction}
In the last two decades, much attention has been paid to the study
of discrete integrable systems, which resulted in the discovery of
second-order integrable mappings, i.e. the Quispel-Robert-Thomson
(QRT) mapping \cite{QRT,QRT2} and discrete Painlev\'e equations
\cite{painleve}. A large number of papers have been devoted to these 
integrable mappings \cite{painleve-review,tsuda}. In contrast, the
examples of higher-dimensional integrable mappings are a few
\cite{HKY,Capel,Iatrou,MT,RQ,Fordy}. It is paramount to find and study
higher-dimensional integrable mappings in order to develop the
theory of discrete integrable systems.

One way to achieve above goal is to construct integrable
discretizations from some known integrable differential equations
(DEs).
Actually, several powerful recipes in this aspect have been
proposed, such as a method based on bilinear form \cite{Hirota77},
and a method based on Lax pair \cite{AL1,AL2,AL3,suris-book}.
Although these discretization methods have produced many discrete
integrable systems, integrable discrete analogues of some
many well-known integrable higher-dimensional (coupled)
Hamiltonian systems are still missing.

In this Letter, we
construct integrable discretizations of a two-dimensional
Hamiltonian system with a quartic potential, whose Hamiltonian can
be expressed as
\begin{equation}
H=\frac{1}{2}(p_1^2+p_2^2)+Aq_1^2+Bq_2^2+\alpha q_1^4 +\beta
q_2^4+\delta q_1^2q_2^2\,.\label{hamilton-cont}
\end{equation}
Thus, the equations of motion are given by
\begin{equation}
\left\{
\begin{array}{l}
\dot{q}_1=\frac{\partial H}{\partial p_1}=p_1\,,\\
\dot{q}_2=\frac{\partial H}{\partial p_2}=p_2\,,\\
\dot{p}_1=-\frac{\partial H}{\partial q_1}=-2Aq_1-2q_1(2\alpha
 q_1^2+\delta q_2^2)\,,\\
\dot{p}_2=-\frac{\partial H}{\partial q_2}=-2Bq_2-2q_2(2\beta
 q_2^2+\delta q_1^2)\,,
\end{array}
\right.
\end{equation}
or
\begin{equation}
\left\{
\begin{array}{l}
\ddot{q}_1=-2Aq_1-2q_1(2\alpha
 q_1^2+\delta q_2^2)\,,\\
\ddot{q}_2=-2Bq_2-2q_2(2\beta
 q_2^2+\delta q_1^2)\,.
\end{array}
\right.
\label{cont-quartic}
\end{equation}
The above Hamiltonian system is known to be integrable for the
following
four cases \cite{hietarinta,RGB,LS}:\\
\,(I)\, $\beta=\alpha$, $\delta=2\alpha$, \quad $\alpha$, $A$, $B$ arbitrary\,,\\
\,(II)\, $\beta=\alpha$, $\delta=6\alpha$, $A=B$, \quad $\alpha$, $A$
arbitrary\,,\\
\,(III)\, $\beta=16\alpha$, $\delta=12\alpha$, $B=4A$, \quad
$\alpha$, $A$ arbitrary\,,\\
\,(IV)\, $\beta=8\alpha$, $\delta=6\alpha$,  $B=4A$, \quad $\alpha$,
$A$ arbitrary\,, \\
with the corresponding second integrals listed as follows: \\
\,\,(I) $I(q_1,q_2,p_1,p_2)=(q_1p_2-q_2p_1)^2+\frac{B-A}{\alpha}\{
p_1^2+2q_1^2[A+\alpha (q_1^2+q_2^2)]
\}$\,,\\
\,\,(II) $I(q_1,q_2,p_1,p_2)=p_1p_2+2q_1q_2[A+2\alpha (q_1^2+q_2^2)]$\,,\\
\,\,(III) $I(q_1,q_2,p_1,p_2)=(q_1p_2-q_2p_1)p_1+2q_1^2q_2
[A+\alpha (q_1^2+2q_2^2)]$\,,\\
\,\,(IV) $I(q_1,q_2,p_1,p_2)=4\alpha q_1^2(q_1p_2-2q_2p_1)^2 +\{
p_1^2+2q_1^2[A+\alpha (q_1^2+2q_2^2)]
\}^2$\,.\\

The cases (I), (II) and (III) are separable in ellipsoidal,
Cartesian and paraboidal coordinates, respectively, while
the case (IV) being separable in the general sense
\cite{Ravoson,Romeiras,Adrian}.
Further, the case (I) is equivalent to a
traveling wave reduction of the integrable coupled nonlinear
Schr\"{o}dinger (cNLS) equation (Manakov system), and it also
corresponds to the Garnier system of two particles
\cite{garnier,wojciekowski}.

The present paper is organized as follows: In section 2, we present two
discretizations of the case (I) based on bilinear form. Then, we
show the integrability of these discretizations by deriving discrete
versions of two first integrals for the original Hamiltonian system
(\ref{hamilton-cont}). It should be noted that what we obtain here
are alternative forms of some already known results
\cite{suris,ohta,suris-book}. In section 3,
we derive an integrable discretization of the
case (II) and show its integrals. The key of the discretization is
the separation of variables. In section 4, we construct an
integrable discretization of the case (III) by virtue of the
separation of variable. The approach used in this case is similar to
the one used for the discretization of the Kepler problem
\cite{Minesaki}. We give some remarks and further topics in section 5.

\section{The discretizations of the case (I) and Integrals}
\subsection{The discretizations of the case (I)}
In this section, we consider the integrable discretization of the case
(I), i.e. $\beta=\alpha$, $\delta=2\alpha$ ($\alpha$, $A$ and $B$:
arbitrary). Upon the dependent variable transformation
\[
 q_i=\frac{g_i}{f},\qquad i=1,2\,,
\]
eq. (\ref{cont-quartic}) can be casted into the bilinear forms
\begin{eqnarray}
&&D_t^2g_1\cdot f +2Afg_1=0\,,\label{bi1}\\
&&-D_t^2f\cdot f+4\alpha g_1^2+4\alpha g_2^2=0\,,\\
&&D_t^2g_2\cdot f +2Bfg_2=0\,,\\
&&-D_t^2f\cdot f+4\alpha g_2^2+4 \alpha g_1^2=0\,.\label{bi4}
\end{eqnarray}
Based upon the discetization of above bilinear forms, we can obtain
two integrable discretizations as follows.

{\bf Discretization (a)}\\
An integrable discretization of eq.(\ref{cont-quartic}) for the case
(I) is
\begin{eqnarray}
&&\frac{{q}_{1,n+1}-2q_{1,n}+q_{1,n-1}}{h^2} =-2Aq_{1,n}\nonumber\\
&&\quad -2\alpha
q_{1,n}\left[q_{1,n}(q_{1,n+1}+q_{1,n-1})+ \frac{1-Ah^2}{1-Bh^2}
q_{2,n} (q_{2,n+1}+q_{2,n-1})\right] \,,\label{disc-quartica1}\\
&&\frac{{q}_{2,n+1}-2q_{2,n}+q_{2,n-1}}{h^2}=-2Bq_{2,n}\nonumber\\
&&\quad -2\alpha q_{2,n}\left[q_{2,n}(q_{2,n+1}+q_{2,n-1})+
\frac{1-Bh^2}{1-Ah^2}q_{1,n} (q_{1,n+1}+q_{1,n-1})\right]\,.
\label{disc-quartica2}
\end{eqnarray}

Upon the dependent variable transformation
\begin{equation}\label{var-tran}
 q_{i,n}=\frac{g_{i,n}}{f_n},\qquad i=1,2\,,
\end{equation}
eqs.(\ref{disc-quartica1})-(\ref{disc-quartica2}) are transformed to
\begin{eqnarray}
&&g_{1,n+1}f_{n-1}+g_{1,n-1}f_{n+1}=
\Gamma_1g_{1,n}f_n\,,\label{bilinear-a1}\\
&&(2-2Ah^2)f_{n+1}f_{n-1}-2\alpha h^2\Gamma_1g_{1,n}^2
-2\alpha \frac{1-Ah^2}{1-Bh^2} h^2\Gamma_2 g_{2,n}^2=\Gamma_1f_n^2\,,
\label{bilinear-a2}\\
&&g_{2,n+1}f_{n-1}+g_{2,n-1}f_{n+1}=
\Gamma_2g_{2,n}f_n\,,\label{bilinear-a3}\\
&&(2-2Bh^2)f_{n+1}f_{n-1}-2\alpha h^2\Gamma_2g_{2,n}^2 -2\alpha
\frac{1-Bh^2}{1-Ah^2}h^2\Gamma_1 g_{1,n}^2=\Gamma_2f_n^2\,,
\label{bilinear-a4}
\end{eqnarray}
which are discretization of bilinear forms (\ref{bi1})-(\ref{bi4}).

Multiplying eq.(\ref{disc-quartica1}) by $(q_{1,n+1}-q_{1,n-1})/2$,
eq.(\ref{disc-quartica2}) by $(q_{2,n+1}-q_{2,n-1})/2$, and summing
them up, we obtain an integral of motion
\begin{eqnarray}
&&H_n= \frac 12 p_{1,n}^2+\frac 12 p_{2,n}^2
+Aq_{1,n+1}q_{1,n}+Bq_{2,n+1}q_{2,n}+\alpha q_{1,n+1}^2q_{1,n}^2\nonumber\\
&&\quad +\alpha q_{2,n+1}^2q_{2,n}^2
+\alpha \left(\frac{1-Ah^2}{1-Bh^2}+\frac{1-Bh^2}{1-Ah^2}\right)
q_{1,n+1}q_{1,n}q_{2,n+1}q_{2,n}\,, \label{disc-hamilton1}
\end{eqnarray}
which is a discrete analogue of the Hamiltonian
(\ref{hamilton-cont}) where
\[
p_{1,n}=\frac{q_{1,n+1}-q_{1,n}}{h}\,,\quad
p_{2,n}=\frac{q_{2,n+1}-q_{2,n}}{h} \,.
\]

{\bf Discretization (b)}\\
Another discretization of eq.(\ref{cont-quartic}), which was firstly
proposed by Suris \cite{suris,suris-book}, is of the form

\begin{equation}\label{disc-quarticb1}
\frac{q_{1,n+1}-2q_{1,n}+q_{1,n-1}}{h^2} =-2A q_{1,n}-2\alpha(
q_{1,n}^2+ q_{2,n}^2) (q_{1,n+1}+q_{1,n-1})\,,
\end{equation}

\begin{equation}\label{disc-quarticb2}
\frac{q_{2,n+1}-2q_{2,n}+q_{2,n-1}}{h^2} =-2B q_{2,n}-2\alpha(
q_{1,n}^2+ q_{2,n}^2)(q_{2,n+1}+q_{2,n-1})\,.
\end{equation}

Upon the same transformation (\ref{var-tran}),
eqs.(\ref{disc-quarticb1})-(\ref{disc-quarticb2}) can be casted into
discrete bilinear forms
\begin{eqnarray}
&&g_{1,n+1}f_{n-1}+g_{1,n-1}f_{n+1}=\Gamma_1g_{1,n}f_n\,,\\
&&\frac{2-2Ah^2}{\Gamma_1} f_{n+1}f_{n-1}-2\alpha h^2 g_{1,n}^2
-2 \alpha h^2 g_{2,n}^2=f_n^2\,,\\
&&g_{2,n+1}f_{n-1}+g_{2,n-1}f_{n+1}=\Gamma_2g_{2,n}f_n\,,\\
&&\frac{2-2Bh^2}{\Gamma_2}f_{n+1}f_{n-1}-2\alpha h^2g_{2,n}^2 -2\alpha
h^2 g_{1,n}^2=f_n^2\,.
\end{eqnarray}

Similarly, it can be shown that discretization (b) admits an
integral of motion
\begin{eqnarray}
&&H_n= \frac 12 p_{1,n}^2+ \frac 12 p_{2,n}^2
+Aq_{1,n+1}q_{1,n}+Bq_{2,n+1}q_{2,n}\nonumber\\
&&\quad +\alpha (q_{1,n+1}^2q_{1,n}^2+ q_{2,n+1}^2q_{2,n}^2 +
q_{2,n+1}^2q_{1,n}^2+  q_{2,n}^2q_{1,n+1}^2) \,,
\label{disc-hamilton2}
\end{eqnarray}
which is a discrete analogue of the Hamiltonian
(\ref{hamilton-cont}).

\subsection{The integrals of discretizations of case (I)}
It was shown by Suris \cite{suris,suris-book} that the discretization (b)
admits two integrals
\begin{eqnarray}
&&I_{1,n}=q_{1,n+1}^2+q_{1,n}^2-2(1-Ah^2)q_{1,n+1}q_{1,n}
+2\alpha h^2 q_{1,n+1}^2q_{1,n}^2\nonumber\\
&&\qquad +\frac{2\alpha (1-A h^2)^2}{(1-Ah^2)^2-(1-Bh^2)^2}
(q_{1,n+1}^2q_{2,n}^2+q_{1,n}^2q_{2,n+1}^2)\nonumber\\
&&\qquad - \frac{4\alpha (1-A h^2)(1-B h^2)}{(1-Ah^2)^2-(1-Bh^2)^2}
q_{1,n+1}q_{2,n+1}q_{1,n}q_{2,n}\,,\\
&&I_{2,n}=q_{2,n+1}^2+q_{2,n}^2-2(1-Bh^2)q_{2,n+1}q_{2,n}
+2\alpha h^2 q_{2,n+1}^2q_{2,n}^2\nonumber\\
&&\qquad +\frac{2\alpha (1-B h^2)^2}{(1-Bh^2)^2-(1-Ah^2)^2}
(q_{2,n+1}^2q_{1,n}^2+q_{2,n}^2q_{1,n+1}^2)\nonumber\\
&&\qquad - \frac{4\alpha (1-A h^2)(1-B h^2)}{(1-Bh^2)^2-(1-Ah^2)^2}
q_{1,n+1}q_{2,n+1}q_{1,n}q_{2,n}\,.
\end{eqnarray}
Note that the sum of above two first integrals corresponds to the
discrete analogue of Hamiltonian (\ref{disc-hamilton2}), the
difference corresponds to the second integral of motion.

Next, we show that discretizations (a) and (b) share the same
integrals. Setting $\Gamma_1=1-Ah^2$ and $\Gamma_2=1-Bh^2$ in the
bilinear equations (\ref{bilinear-a1})--(\ref{bilinear-a3}), we
have
\begin{eqnarray}
&& (1-Ah^2)q_{1,n}(q_{2,n+1}+q_{2,n-1})
=(1-Bh^2)q_{2,n}(q_{1,n+1}+q_{1,n-1})\,. \label{relation-a-case1}
\end{eqnarray}
Employing (\ref{relation-a-case1}), it is easily shown that
eqs.(\ref{disc-quartica1})-(\ref{disc-quartica2}) are equivalent to
eqs.(\ref{disc-quarticb1})-(\ref{disc-quarticb2}). Thus, discretization
(a) has the same integrals as discretization (b).

{\bf Special case ($A=B$): }\\
Under this case, the discretization is
\begin{eqnarray}
&&{q}_{1,n+1}+q_{1,n-1}
-2(1-Ah^2)q_{1,n}
+2\alpha h^2(q_{1,n}^2+q_{2,n}^2)(q_{1,n+1}+q_{1,n-1})
=0\nonumber\\
&&\label{disc-quartic-case-I-1-b-special}\\
&&
{q}_{2,n+1}+q_{2,n-1}
-2(1-Ah^2)q_{2,n}
+2\alpha h^2(q_{1,n}^2+q_{2,n}^2)(q_{2,n+1}+q_{2,n-1})
=0\,.
\nonumber\\
&&\label{disc-quartic-case-I-2-b-special}
\end{eqnarray}
Multiplying eq.(\ref{disc-quartic-case-I-1-b-special}) by
$(q_{2,n+1}+q_{2,n-1})$, eq.(\ref{disc-quartic-case-I-2-b-special})
by $(q_{1,n+1}+q_{1,n-1})$, and subtracting them, we arrive at
\begin{eqnarray*}
&&-2(1-Ah^2)q_{1,n}(q_{2,n+1}+q_{2,n-1})
+2(1-Ah^2)q_{2,n}(q_{1,n+1}+q_{1,n-1})=0\,,
\end{eqnarray*}
from which, we have the following second integral
\[
I=q_{1,n}q_{2,n+1}-q_{1,n+1}q_{2,n}\,.
\]

It is known that the case (I) can be obtained by a stationary reduction
of a Manakov system. This fact remains true for the discretization,
i.e. the integrable discretization of the case (I) can be obtained from
a semi-discrete cNLS equation:
\begin{equation}
\left\{
\begin{array}{l}
{\rm i}\frac{\partial}{\partial t}q_{1,n}
+\frac{q_{1,n+1}-2q_{1,n}+q_{1,n-1}}{h^2} \\
\quad =-2A q_{1,n}-(2\alpha
|q_{1,n}|^2+2\alpha |q_{2,n}|^2) (q_{1,n+1}+q_{1,n-1})\,,
\\
{\rm i}\frac{\partial}{\partial t}q_{2,n}+
\frac{q_{2,n+1}-2q_{2,n}+q_{2,n-1}}{h^2} \\
\quad =-2B q_{2,n}-(2\alpha
|q_{1,n}|^2+2\alpha |q_{2,n}|^2)(q_{2,n+1}+q_{2,n-1})\,,
\end{array}\right.
\label{dcnls2}
\end{equation}
by a stationary reduction. The semi-discrete cNLS
equation (\ref{dcnls2}) was studied by Ohta and $N$-soliton
solutions were given in the form of Pfaffian \cite{ohta}.

\section{Integrable discretizations of case (II)}
Let us consider how to build up integrable discretization of the case
(II) ($\beta=\alpha$, $\delta=6\alpha$, $A=B$), where the equations
of motion are given by
\begin{eqnarray}
&&\ddot{q}_1=-2Aq_1-2q_1(2\alpha
 q_1^2+6\alpha q_2^2)\,,\nonumber\\
&&\ddot{q}_2=-2Aq_2-2q_2(2\alpha
 q_2^2+6\alpha q_1^2)\,.\label{case2-cont}
\end{eqnarray}
First, we comment that the method of discretization based on
bilinear form is not able to construct integrable discretization of
the case (II). This is a result of breaking solution structure of
the case (II) by the transformation to the bilinear forms.

Fortunately, an integrable
discretization can be constructed by noting that
eqs.(\ref{case2-cont}) are separable in Cartesian coordinates. It is
known that eqs.(\ref{case2-cont}) can be decoupled to two anharmonic
oscillators by a simple transformation: $u_{1}=q_{1}+q_{2}$,
$u_{2}=q_{1}-q_{2}$:
\begin{eqnarray*}
&&\ddot{u}_1=-2Au_1-4\alpha
 u_1^3\,,\nonumber\\
&&\ddot{u}_2=-2Au_2-4\alpha
 u_2^3\,. \label{osillators-decoupled}
\end{eqnarray*}

An integrable discretization
\begin{eqnarray}
&&\frac{u_{1,n+1}-2u_{1,n}+u_{1,n-1}}{h^2} =-2A u_{1,n}-4\alpha
u_{1,n}^2 \frac{u_{1,n+1}+u_{1,n-1}}{2}\,,
\nonumber \\
&&\frac{u_{2,n+1}-2u_{2,n}+u_{2,n-1}}{h^2} =-2A u_{2,n}-4\alpha
u_{2,n}^2\frac{u_{2,n+1}+u_{2,n-1}}{2}\,, \label{decoupled_dis}
\end{eqnarray}
is known, then we can easily give two integrals which are
discrete Hamiltonians of (\ref{osillators-decoupled}) as follows:
\begin{eqnarray}
&&H_1= \frac{1}{2} p_{u_{1},n}^2 + Au_{1,n}u_{1,n+1} +\alpha
u_{1,n}^2 u_{1,n+1}^2\,,
\nonumber \\
&&H_2= \frac{1}{2} p^2_{u_{2},n} + Au_{2,n}u_{2,n+1} +\alpha
u_{2,n}^2 u_{2,n+1}^2, \label{Hamiltonian-dis}
\end{eqnarray}

Substituting $u_{1,n}=q_{1,n}+q_{2,n}$ and $u_{2,n}=q_{1,n}-q_{2,n}$
into two discrete anharmonic oscillators ({\ref{decoupled_dis}), we
have the following discretization of the case (II) (symmetric
discretization):
\begin{eqnarray}
&&\frac{q_{1,n+1}-2q_{1,n}+q_{1,n-1}}{h^2} \nonumber\\
&&\quad =-2A q_{1,n}-(\alpha q_{1,n}^2+\alpha q_{2,n}^2)
\frac{q_{1,n+1}+q_{1,n-1}}{2} -2\alpha q_{1,n}q_{2,n}
\frac{q_{2,n+1}+q_{2,n-1}}{2}\,,
\nonumber \\
&&\frac{q_{2,n+1}-2q_{2,n}+q_{2,n-1}}{h^2} \nonumber \\
&&\quad =-2A q_{2,n}-(\alpha q_{1,n}^2+\alpha
q_{2,n}^2)\frac{q_{2,n+1}+q_{2,n-1}}{2} -2\alpha
q_{2,n}q_{1,n}\frac{q_{1,n+1}+q_{1,n-1}}{2}\,.\nonumber
\\
&& \quad \label{discrete-case2}
\end{eqnarray}

Similarly, we can obtain an asymmetric discretization of the case (II)
by substituting $u_{1,n}=q_{1,n}+q_{2,n}$ and
$u_{2,n-1}=q_{1,n}-q_{2,n}$ into two discrete anharmonic oscillators
(\ref{decoupled_dis})
\begin{eqnarray}
&&\frac{q_{1,n+1}-2q_{1,n}+q_{1,n-1}}{h^2} \nonumber\\
&&\quad =-2A q_{1,n}-(\alpha q_{1,n}^2+\alpha q_{2,n-1}^2)
\frac{q_{1,n+1}+q_{1,n-1}}{2}
-2\alpha q_{1,n}q_{2,n-1}\frac{q_{2,n}+q_{2,n-2}}{2}\,,
\nonumber \\
&&\frac{q_{2,n+1}-2q_{2,n}+q_{2,n-1}}{h^2} \nonumber \\
&&\quad =-2A q_{2,n}-(\alpha q_{1,n+1}^2+\alpha
q_{2,n}^2)\frac{q_{2,n+1}+q_{2,n-1}}{2} -2\alpha
q_{2,n}q_{1,n+1}\frac{q_{1,n+2}+q_{1,n}}{2}\,.\nonumber
\\
&& \quad \label{discrete-case2-2}
\end{eqnarray}

It can be easily shown that $(H_1+H_2)/2$ and $(H_1-H_2)/2$ in
(\ref{Hamiltonian-dis}) are discrete versions of two integrals for
the original Hamiltonian system (\ref{hamilton-cont})
\begin{eqnarray}
&&I_1= \frac{1}{2} p^2_{1,n}+ \frac{1}{2} p^2_{2,n} +
A(q_{1,n}q_{1,n+1}+q_{2,n}q_{2,n+1}) \nonumber
\\ &&\quad  +\alpha
(q^2_{1,n}q^2_{1,n+1}+q^2_{2,n}q^2_{2,n+1}+4 q_{1,n}q_{1,n+1}
q_{2,n}q_{2,n+1})\,,
\nonumber \\
&&I_2= p_{1,n}p_{2,n} +
A(q_{1,n}q_{2,n+1}+q_{2,n}q_{1,n+1}) \nonumber \\
&&\quad +2\alpha (q_{1,n}q_{2,n+1}+q_{2,n}q_{1,n+1})
(q_{1,n}q_{1,n+1}+q_{2,n}q_{2,n+1})\,,
\end{eqnarray}
by the dependent variable transformation $u_{1,n}=q_{1,n}+q_{2,n}$ and
$u_{2,n-1}=q_{1,n}-q_{2,n}$.

In Ref.\cite{HKY}, the more general form of integrable
discretization of the anharmonic oscillator
\[
 \frac{d^2x}{dt^2}+ax+bx^3=0\,,
\]
was presented.
The form of the discretization of the anharmonic oscillator is
\begin{eqnarray*}
&&\frac{u_{n+1}-2u_n+u_{n-1}}{h^2}+2A[c_{11}u_n+c_{12}(u_{n+1}+u_{n-1})]\\
&&\quad \quad \quad +4\alpha
[c_{21}u_{n+1}u_{n}u_{n-1}+c_{22}u_n^2(u_{n+1}+u_{n-1})]=0 \,.
\end{eqnarray*}
Using this discretization and the transformation
$u_{1,n}=q_{1,n}+q_{2,n}$, $u_{2,n}=q_{1,n}-q_{2,n}$, we can make a
more general integrable symmetric discretization of the case (II),
\begin{eqnarray}
&& \frac{q_{1,n+1}-2q_{1,n}+q_{1,n-1}}{h^2} =-2A
\left(c_{11}q_{1,n}+c_{12}(q_{1,n+1}+q_{1,n-1})\right) \nonumber \\
&&\quad  -4\alpha c_{22} (q_{1,n}^2+2\alpha q_{2,n}^2)
(q_{1,n+1}+q_{1,n-1})  -8\alpha
c_{22}q_{1,n}(q_{2,n}(q_{2,n+1}+q_{2,n-1})) \nonumber \\
&&\quad  -2\alpha
c_{21} \left[ q_{2,n} q_{2,n+1} (q_{1,n+1}+q_{1,n-1}) + q_{1,n}
q_{1,n+1}q_{1,n-1} + q_{2,n} q_{2,n+1} q_{2,n-1}\right]\,,
\nonumber \\
&& \frac{q_{2,n+1}-2q_{2,n}+q_{2,n-1}}{h^2} =-2A
\left(c_{11}q_{2,n}+c_{12}(q_{2,n+1}+q_{2,n-1})\right) \nonumber \\
&& \quad -4\alpha c_{22} (q_{2,n}^2+2\alpha q_{1,n}^2)
(q_{2,n+1}+q_{2,n-1})  -8\alpha
c_{22}q_{2,n}(q_{1,n}(q_{1,n+1}+q_{1,n-1})) \nonumber \\
&&\quad  -2\alpha
c_{21} \left[ q_{1,n} q_{1,n+1} (q_{2,n+1}+q_{2,n-1}) + q_{1,n}
q_{1,n+1}q_{1,n-1} + q_{2,n} q_{2,n+1} q_{2,n-1}\right]\,.\nonumber
\\
&& \quad \label{discrete-case3}
\end{eqnarray}
The constants $c_{11}$, $c_{12}$, $c_{21}$
and $c_{22}$ are given in Ref.\cite{HKY}, i.e.
\[
c_{11}=\gamma_{11}\,,\quad  c_{12}=\gamma_{12}\,,\quad
c_{21}=(1-a\gamma_{11}/2)\gamma_{21}\,,\quad
c_{22}=(1+a\gamma_{12})\gamma_{22}/2\,,
\]
while $\gamma_{11}$,$\gamma_{12}$,$\gamma_{21}$,$\gamma_{22}$
satisfying the relations $\gamma_{11}+2\gamma_{12}=1$,
$\gamma_{21}+\gamma_{22}=1$.

 From the result in Ref.\cite{HKY}, we
can construct integrals of eq.(\ref{discrete-case3}):
\begin{eqnarray*}
&&I_1=H_1+H_2\,,\\
&&I_2=H_1-H_2\,,
\end{eqnarray*}
where
\begin{eqnarray}
&&H_1=\bigl(c_{11}[c_{12}((q_{1,n}+q_{2,n})^2+(q_{1,n-1}+q_{2,n-1})^2)
-c_{11}(q_{1,n}+q_{2,n})(q_{1,n-1}+q_{2,n-1})]\nonumber\\
&&\quad \quad \quad \quad +2\alpha (c_{11}c_{22}+c_{11}c_{21})
(q_{1,n}+q_{2,n})^2(q_{1,n-1}+q_{2,n-1})^2\bigr)\nonumber\\
&&\quad
/\bigl(c_{11}[c_{11}+2\alpha
c_{21}((q_{1,n}+q_{2,n})^2+(q_{1,n-1}+q_{2,n-1})^2)]\nonumber\\
&&\quad \quad \quad
+4\alpha^2c_{21}^2(q_{1,n}+q_{2,n})^2(q_{1,n-1}+q_{2,n-1})^2\bigr)\,,\nonumber
\\
&&H_2=\bigl(c_{11}[c_{12}((q_{1,n}-q_{2,n})^2+(q_{1,n-1}-q_{2,n-1})^2)
-c_{11}(q_{1,n}-q_{2,n})(q_{1,n-1}-q_{2,n-1})]\nonumber\\
&&\qquad \qquad  +2\alpha (c_{11}c_{22}+c_{11}c_{21})
(q_{1,n}-q_{2,n})^2(q_{1,n-1}-q_{2,n-1})^2\bigr)\nonumber\\
&&\quad /\bigl(c_{11}[c_{11}+2\alpha c_{21}
((q_{1,n}-q_{2,n})^2+(q_{1,n-1}-q_{2,n-1})^2)]\nonumber \\
&&\quad \quad \quad
+4\alpha^2 c_{21}^2(q_{1,n}-q_{2,n})^2(q_{1,n-1}-q_{2,n-1})^2\bigr)\,.
\nonumber
\end{eqnarray}

\section{Integrable discretization of case (III)}

Let us consider the case (III) ($\beta=16\alpha$, $\delta=12\alpha$,
$B=4A$), whose Hamiltonian is
\begin{equation}
H=\frac{1}{2}(p_1^2+p_2^2)+Aq_1^2+4Aq_2^2+\alpha q_1^4 +16
\alpha q_2^4+12\alpha q_1^2q_2^2\,.\label{hamilton-cont-case3}
\end{equation}
The equations of motion are therefore given by
\begin{equation}
\left\{
\begin{array}{l}
\ddot{q}_1=-2Aq_1-2q_1(2\alpha
 q_1^2+12\alpha q_2^2)\,,\\
\ddot{q}_2=-8Aq_2-2q_2(32\alpha
 q_2^2+12\alpha q_1^2)\,.
\end{array}
\right.\label{case3-cont}
\end{equation}
Eqs.(\ref{case3-cont}) are also separable. Upon the canonical
transformation called the Levi-Civita transformation
\begin{equation}
 q_1=u_1u_2\,,\quad q_2=\frac{1}{2}(u_1^2-u_2^2)\,,\quad
p_1=\frac{u_1p_{u2}+u_2p_{u1}}{u_1^2+u_2^2}\,,\quad
p_2=\frac{u_1p_{u2}-u_2p_{u1}}{u_1^2+u_2^2}\,,\label{levi}
\end{equation}
eq.(\ref{hamilton-cont-case3}) is converted to
\begin{equation}
H(p_{u1},p_{u2},u_1,u_2)
=\frac{1}{u_1^2+u_2^2}
\left(\frac{1}{2}(p_{u1}^2+p_{u2}^2)+A(u_1^6+u_2^6)
+\alpha (u_1^{10}+u_2^{10})\right)\,
\,.\label{hamilton-case3-trans}
\end{equation}
By taking the level set $\lambda$ of Hamiltonian, i.e.
\[
H=\lambda\,,
\]
and introducing a new variable $s$ defined by
\[
 dt=(u_1^2+u_2^2)ds\,,
\]
we obtain
\begin{eqnarray}
\frac{du_k}{ds}=p_{uk}\,,\quad \frac{dp_{uk}}{ds}
=-2\lambda u_k+6Au_k^5+10\alpha
 u_k^9\,,
\quad k=1,2\,,\label{case3-separation}
\end{eqnarray}
or
\begin{equation}
\left\{
\begin{array}{l}
\frac{d^2 u_1}{ds^2}=-2\lambda u_1+6Au_1^5+10\alpha u_1^9\,, \\
\frac{d^2 u_2}{ds^2}=-2\lambda u_2+6Au_2^5+10\alpha u_2^9\,.
\end{array}
\right.
\end{equation}
The discretization of eq.(\ref{case3-separation}) is given as follows:
\begin{eqnarray}
&& \frac{u_{k,n+1}-u_{k,n}}{s_{(n+1)}-s_{(n)}}
=\frac{p_{uk,n+1}+p_{uk,n}}{2}\,,\nonumber
\\
&&\frac{p_{uk,n+1}-p_{uk,n}}{s_{(n+1)}-s_{(n)}}
= -2\lambda u_{k,n} + 2A u_{k,n}^3
\frac{u_{k,n+1}^3-u_{k,n-1}^3}{u_{k,n+1}-u_{k,n-1}}
+ 2\alpha u_{1,n}^5
\frac{u_{k,n+1}^5-u_{k,n-1}^5}{u_{k,n+1}-u_{k,n-1}}\,,\nonumber\\
&& \qquad \qquad  k=1,2\,,\label{case3-discrete}
\end{eqnarray}
where
\[
 p_{u_k,n}=\frac{u_{k,n+1}-u_{k,n}}{s_{(n+1)}-s_{(n)}}\,.
\]

It is shown that the integrals are
\[
I_k = \frac{1}{2} p_{uk,n}^2 + \frac{\lambda}{2} u_{k,n+1}u_{k,n} +A
u_{k,n+1}^3u_{k,n}^3+ u_{k,n+1}^5u_{k,n}^5\,, \qquad \qquad  k=1,2.
\]

Introducing a discrete variable $t_n$ as in \cite{Minesaki}:
\begin{equation}
 t_{n+1}-t_n=(u_{1,n}^2+u_{2,n}^2)(s_{n+1}-s_n)\,, \label{d-var-change}
\end{equation}
eq.(\ref{case3-discrete}) can be rewritten as
\begin{eqnarray}
&& \frac{u_{k,n+1}-u_{k,n}}{t_{(n+1)}-t_{(n)}}
=\frac{p_{uk,n+1}+p_{uk,n}}{2(u_{1,n}^2+u_{2,n}^2)}\,,\nonumber
\\
&&\frac{p_{uk,n+1}-p_{uk,n}}{t_{(n+1)}-t_{(n)}}\nonumber\\
&& \quad =\frac{1}{(u_{1,n}^2+u_{2,n}^2)}
\left( -2\lambda u_{k,n} + 2A u_{k,n}^3
\frac{u_{k,n+1}^3-u_{k,n-1}^3}{u_{k,n+1}-u_{k,n-1}}
+ 2\alpha u_{1,n}^5
\frac{u_{k,n+1}^5-u_{k,n-1}^5}{u_{k,n+1}-u_{k,n-1}}\right)\,,\nonumber\\
&& \qquad \qquad  k=1,2\,.\label{case3-discrete2}
\end{eqnarray}
Since
\[
\lambda=H\equiv
\frac{1}{u_1^2+u_2^2}
\left(\frac{1}{2}(p_{u1}^2+p_{u2}^2)+A(u_1^6+u_2^6)
+\alpha (u_1^{10}+u_2^{10})\right)\,,
\]
we have
\begin{eqnarray}
&& \frac{u_{k,n+1}-u_{k,n}}{t_{(n+1)}-t_{(n)}}
=\frac{p_{uk,n+1}+p_{uk,n}}{2(u_{1,n}^2+u_{2,n}^2)}\,,\nonumber
\\
&&\frac{p_{uk,n+1}-p_{uk,n}}{t_{(n+1)}-t_{(n)}}\nonumber\\
&&\quad
=\frac{1}{(u_{1,n}^2+u_{2,n}^2)}
\Bigr(
\frac{-2}{u_{1,n}^2+u_{2,n}^2}\nonumber\\
&&\times
\left(\frac{1}{2}(p_{u1,n}^2+p_{u2,n}^2)+A(u_{1,n}^6+u_{2,n}^6)
+\alpha (u_{1,n}^{10}+u_{2,n}^{10})
\right) u_{k,n} \nonumber\\
&&\quad \quad  + 2A u_{k,n}^3
\frac{u_{k,n+1}^3-u_{k,n-1}^3}{u_{k,n+1}-u_{k,n-1}}
+ 2\alpha u_{1,n}^5
\frac{u_{k,n+1}^5-u_{k,n-1}^5}{u_{k,n+1}-u_{k,n-1}}\Bigr)\,,\nonumber\\
&& \qquad \qquad  k=1,2\,.\label{case3-discrete3}
\end{eqnarray}
Using the Levi-Civita transformation (\ref{levi}),
we can compute values of $q_1,q_2, p_1, p_2$.

\section{Conclusions}

In this paper we have searched for integrable discretization of a
two-dimensional Hamiltonian system with a quartic potential. Among
the four integrable cases, we have succeeded in constructing
integrable discretizations for the first three cases, i.e., (I),
(II) and (III). The key of integrable discretization is the
separation of variables.

Since our method of discretization using separation of variables can
be applied to many higher-dimensional Hamiltonian systems, it is
possible to obtain various integrable discretizations of
higher-dimensional Hamiltonian systems. These aspects will be
discussed in the forthcoming paper.

\section*{Acknowledgments}
B.F. is grateful for the support of the U.S. Army Research Office
under Contract No. W911NF-05-1-0029. K.M. acknowledges support from
the 21st Century COE program ``Development of Dynamic Mathematics
with High Functionality'' at the Faculty of Mathematics, Kyushu
University and the Ministry of Education, Science, Sports and
Culture, Grant-in-Aid for Young Scientists. 

\end{document}